%% file: th.tex
\documentstyle[sprocl,epsfig]{article}

\def\be{\begin{equation}}
\def\ee{\end{equation}}
\def\bea{\begin{eqnarray}}
\def\eea{\end{eqnarray}}

\begin{document}
\input{main}

\end{document}

%% file: main.tex
\title{TRANSITION FROM PHOTOPRODUCTION TO DEEP-INELASTIC SCATTERING IN
  JETPRODUCTION AT HERA IN NLO QCD}

\author{B. P\"OTTER}
\address{II. Institut f\"ur Theoretische Physik, Universit\"at Hamburg\\ 
  Luruper Chaussee 149, D-22761 Hamburg, Germany
  \\ e-mail: poetter@mail.desy.de}

\maketitle\abstracts{We present results from NLO QCD calculations
on the production of jets in $ep$ collisions at HERA in a continuous
range of photon virtuality $Q^2$. Special attention is given to the
collinear singularities from the virtual photon and the resolved
virtual photon component of the cross sections. Comparisons with dijet
data from H1 are shown and the infrared sensitivity of dijet cross
sections is discussed.}

\section{Introduction}

High $E_T$ jet production in $ep$ scattering experiments is a very
important field to study QCD. The electron interacts with the proton
via a photon with virtuality $Q^2$. Two regions of photon virtuality
have been studied now for quite some time, one being the region of
nearly on-shell photons $Q^2\simeq 0$ GeV$^2$ (photoproduction
regime), the other being the region of large $Q^2$ (deep-inelastic
regime). Perturbative QCD calculations exist in NLO for
photoproduction\cite{1,2} and deep-inelastic scattering\cite{3,4,5}
and the comparison of data and theory gives satisfying results.

It is well known that in photoproduction the photon can either interact
directly or via its parton content with the
partons from the proton, which gives the {\it direct} and 
{\it resolved} components of the cross section. In deep-inelastic
scattering the resolved component is believed to be negligible. 
The question arises, up to which virtualities the
resolved virtual photon component needs to be considered in the jet
cross sections. Only limited data exist for extracting information on
the parton content of the virtual photon. Two LO parametrizations of
the virtual photon structure function are available.\cite{6,7} Some
calculations of jet production with resolved virtual photons exist in
LO.\cite{8} However, LO calculations show a large scale and scheme
dependence and do not depend on any jet algorithm. This situation is
improved appreciably by doing NLO calculations. These have been
performed recently taking care of the initial state singularities of
the virtual photon.\cite{9,10} Here we present a computer program
based on these calculations, in which the direct and resolved virtual
photon components  are implemented in NLO QCD for calculating jet
cross sections in $ep$ collisions. 

\section{Jetproduction in a Continuous Range of Photon Virtuality}

The deep inelastic scattering (DIS) cross section normally does not contain
a resolved component. We will label the component with the direct
coupling of the photon to the subprocess as DIR. It is available
in NLO QCD.\cite{5} The DIR cross section contains
terms of the type $-P_{q\leftarrow \gamma} \ln (Q^2/E_T^2)$ 
that become large for $Q^2\ll E_T^2$ and spoil the convergence of the
perturbative expansion. They have to be subtracted from the DIR component
and absorbed into the parton distribution function (PDF) of the
virtual photon.\cite{9} The subtraction is done such that the
photoproduction limit is obtained by taking $Q^2\to 0$. The full cross
section is then a sum (SUM) of the resolved (RES), available in NLO
QCD,\cite{1} and the subtracted direct (DIRS) cross sections. 
The components DIR, DIRS and RES can be calculated with the NLO 
program {\tt JetViP}.

We show the limiting cases of large and small $Q^2$ in the
following. All calculations presented in this paper are done in the
hadronic cms, except those for photoproduction, which are done in the
laboratory frame. 
Fig.\ 1 a and b show the single-jet inclusive cross section
integrated over the whole kinematically allowed $\eta$-range with
$y\in [0.1,0.6]$ as a function of $E_T$ in LO and 
NLO for the two $Q^2$-ranges $Q^2\in [0.5,1]$ GeV$^2$ and $Q^2\in
[500,1000]$ GeV$^2$ with the scale $\mu^2=Q^2$. The predictions by
{\tt JetViP} (lines) are compared with those made by DISENT \cite{4}
(histograms) for the purely deep-inelastic scattering component DIR.
The agreement between the two programs is very good in both
$Q^2$ bins. As can be seen, the NLO correction lies a factor of $4$
above the LO curve for the smaller $Q^2$ bin. This is in contrast to
the deep inelastic region shown in Fig.\ 1 b, where the NLO curve is
only 25\% above the LO curve. The large correction for the small
$Q^2$-region stems from the large logarithm described above and
indicates the necessity to introduce a resolved photon component. The
other limiting case, photoproduction, is shown in Fig.~2 for inclusive
single-jet cross sections for $y\in [0.05,0.6]$ with the scale
$\mu^2=E_T^2$. The left plot in Fig.~2 shows the $E_T$ spectrum,
integrated over the kinematically allowed $\eta$ region for 
the subtracted direct DIRS (dashed) and resolved (full) cross
sections in NLO QCD, compared to the predictions from a program by
M.~Klasen for photoproduction (dots).\cite{1} The agreement is perfect
for both components. This also holds for the right plot in Fig.~2,
which shows the $\eta$-distribution integrated over $E_T>5$ GeV. We
conclude that {\tt JetViP} reproduces the limiting cases of DIS and
photoproduction very well. 

\section{Transition from Low to High $Q^2$}

We have calculated single- and dijet cross sections for the
intermediate $Q^2$ bins shown in Tab.\ 1.\cite{10} We use
the scale $\mu^2=Q^2+E_T^2$ in the transition region. 

\begin{figure}[hhh]
  \unitlength1mm
  \begin{picture}(80,80)
    \put(0,11){\epsfig{file=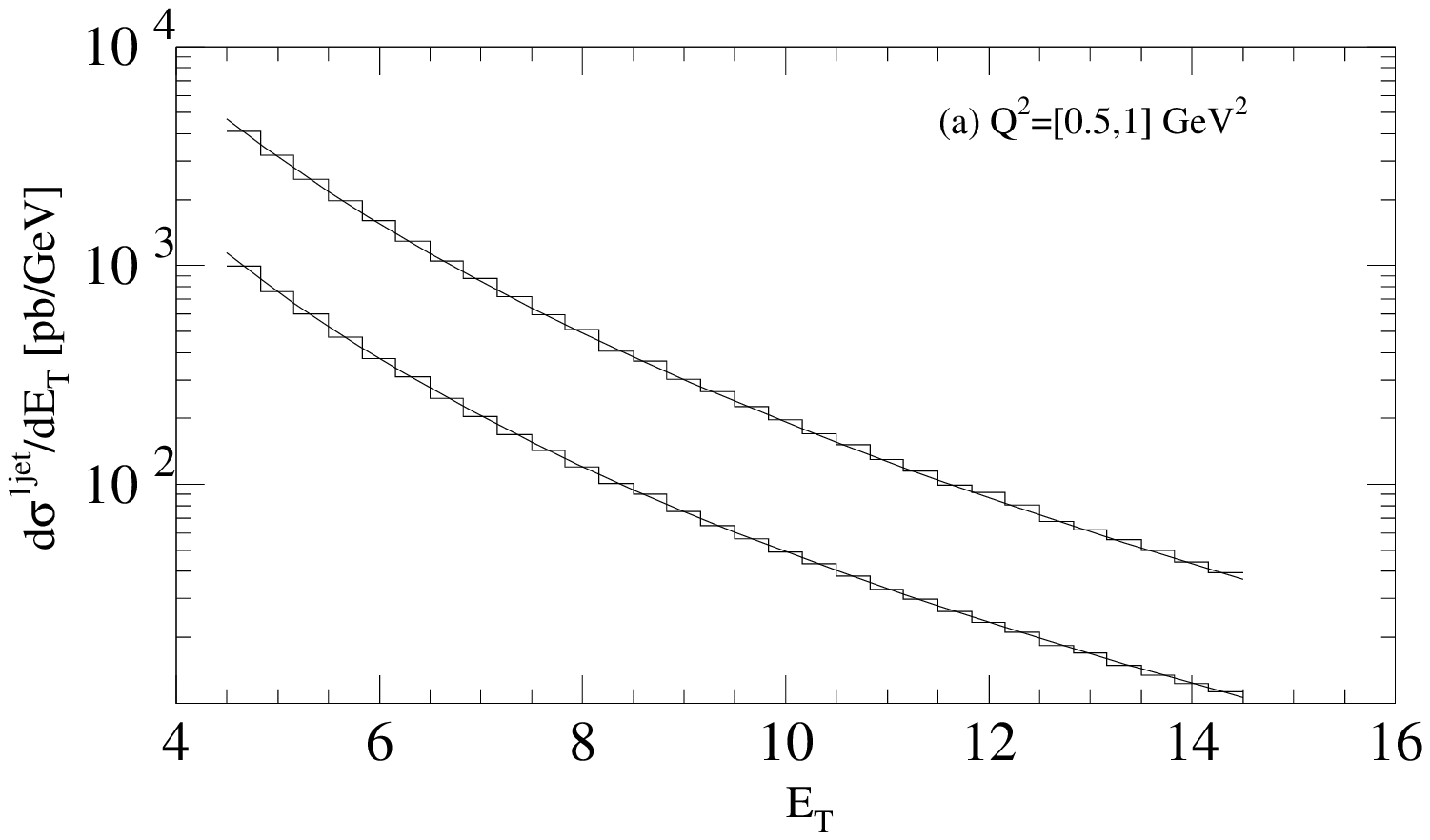,width=6.5cm,height=8cm}}
    \put(60,11){\epsfig{file=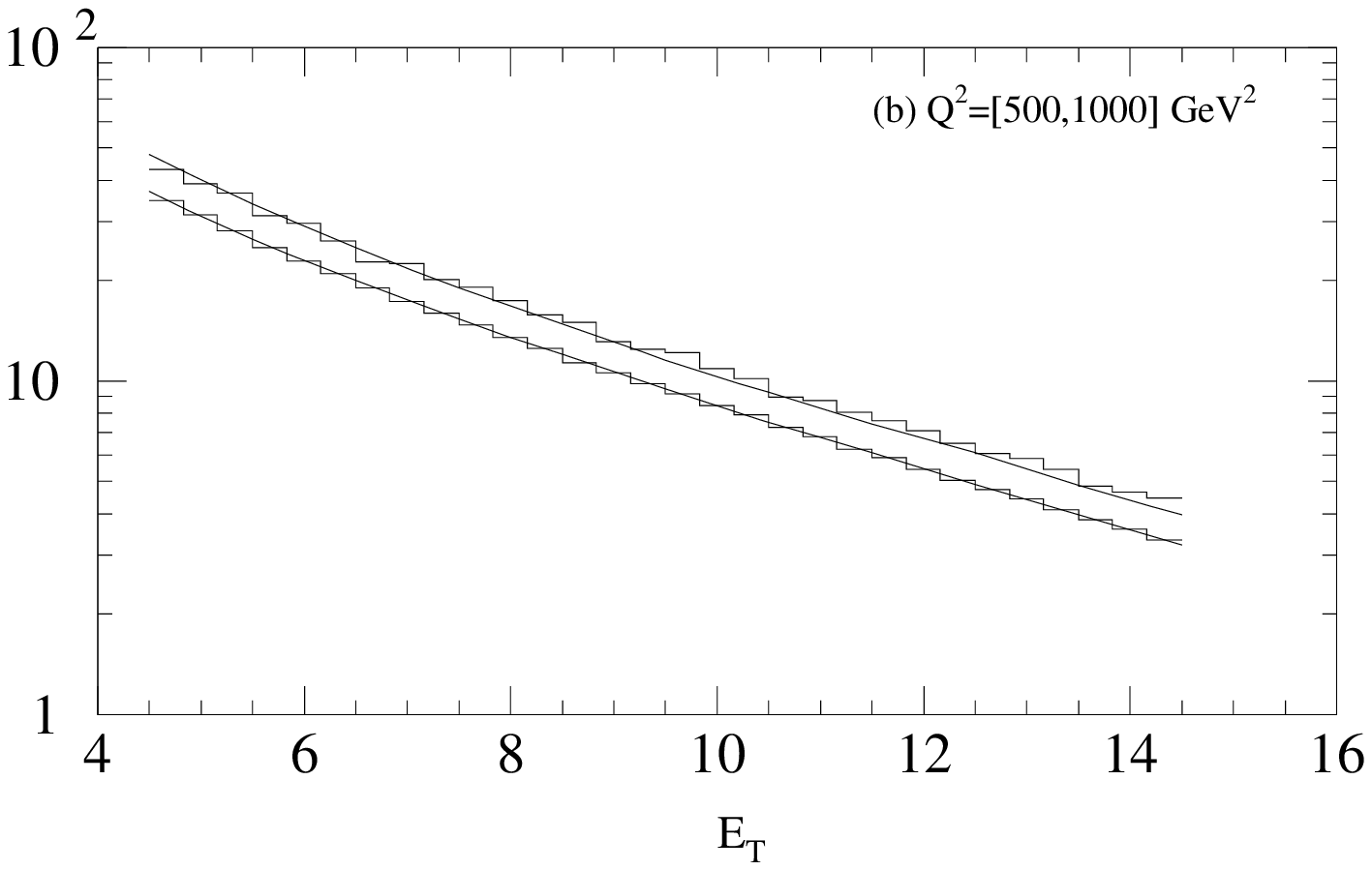,width=6.5cm,height=8cm}}
    \put(0,-37){\epsfig{file=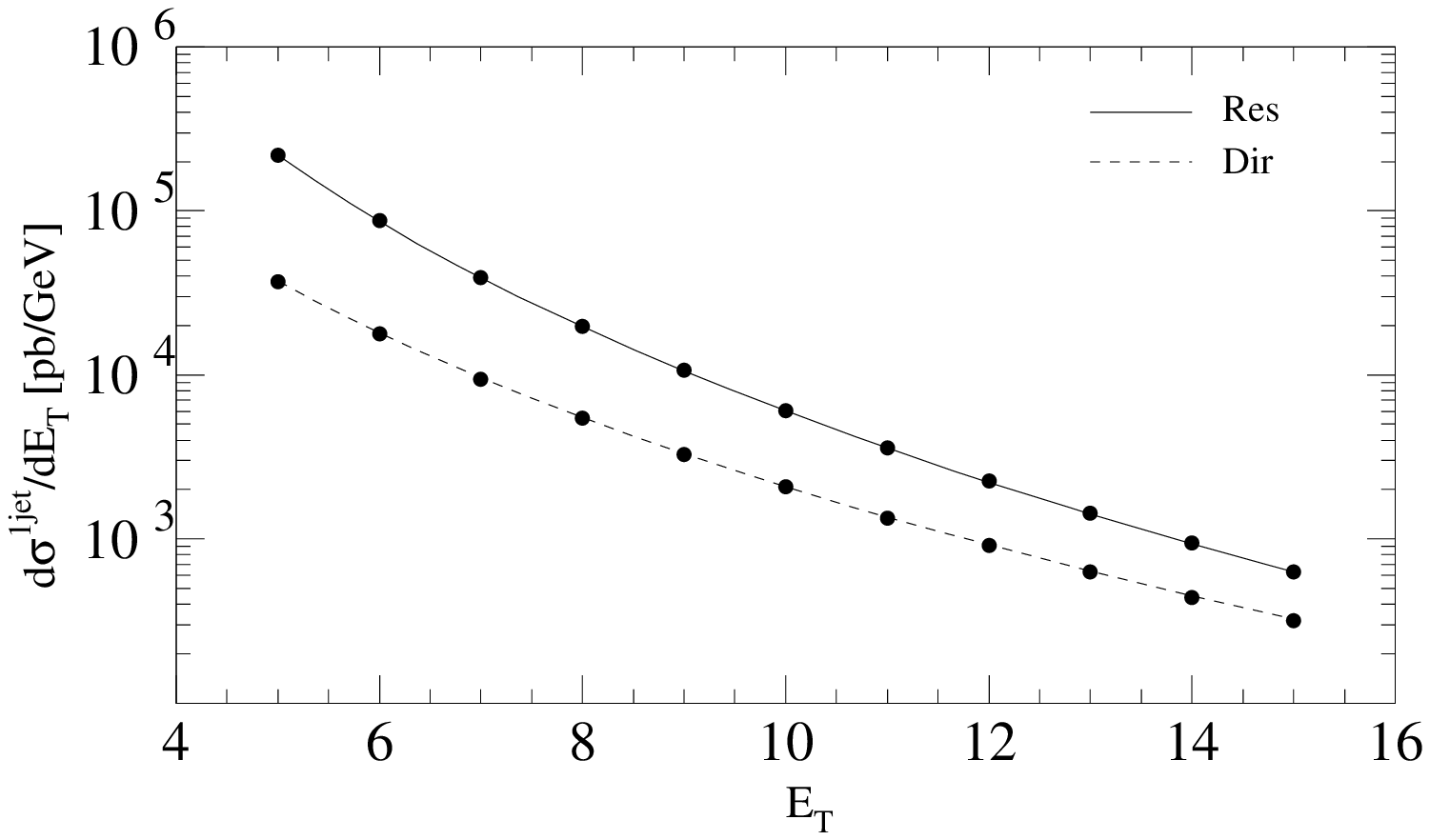,width=6.5cm,height=8cm}}
    \put(0,45){\parbox[t]{4.7in}{\sloppy Figure 1 a,b: Inclusive
        single-jet cross sections in DIS in LO (lower) and NLO (upper)
        compared to DISENT \cite{4} (histograms) for two different
        $Q^2$-bins.}}
    \put(60,-37){\epsfig{file=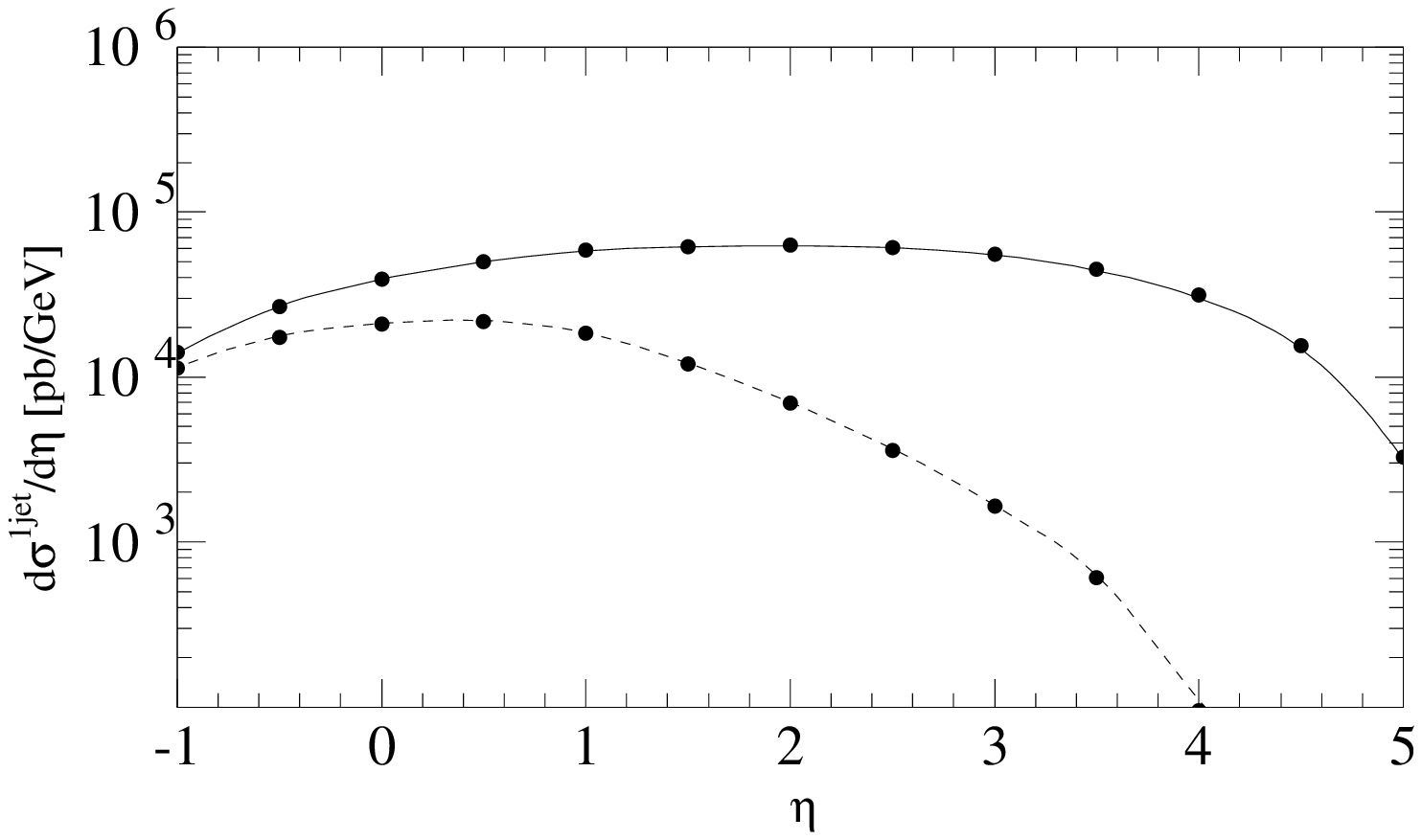,width=6.5cm,height=8cm}}
    \put(0,-2){\parbox[t]{4.7in}{\sloppy Figure 2: Inclusive
        single-jet cross sections in photoproduction for the
        direct (dashed) and resolved (full) components compared to
        Klasen/Kramer \cite{1} (dots).}}
  \end{picture}
  \begin{picture}(80,80)
    \put(0,-2){\epsfig{file=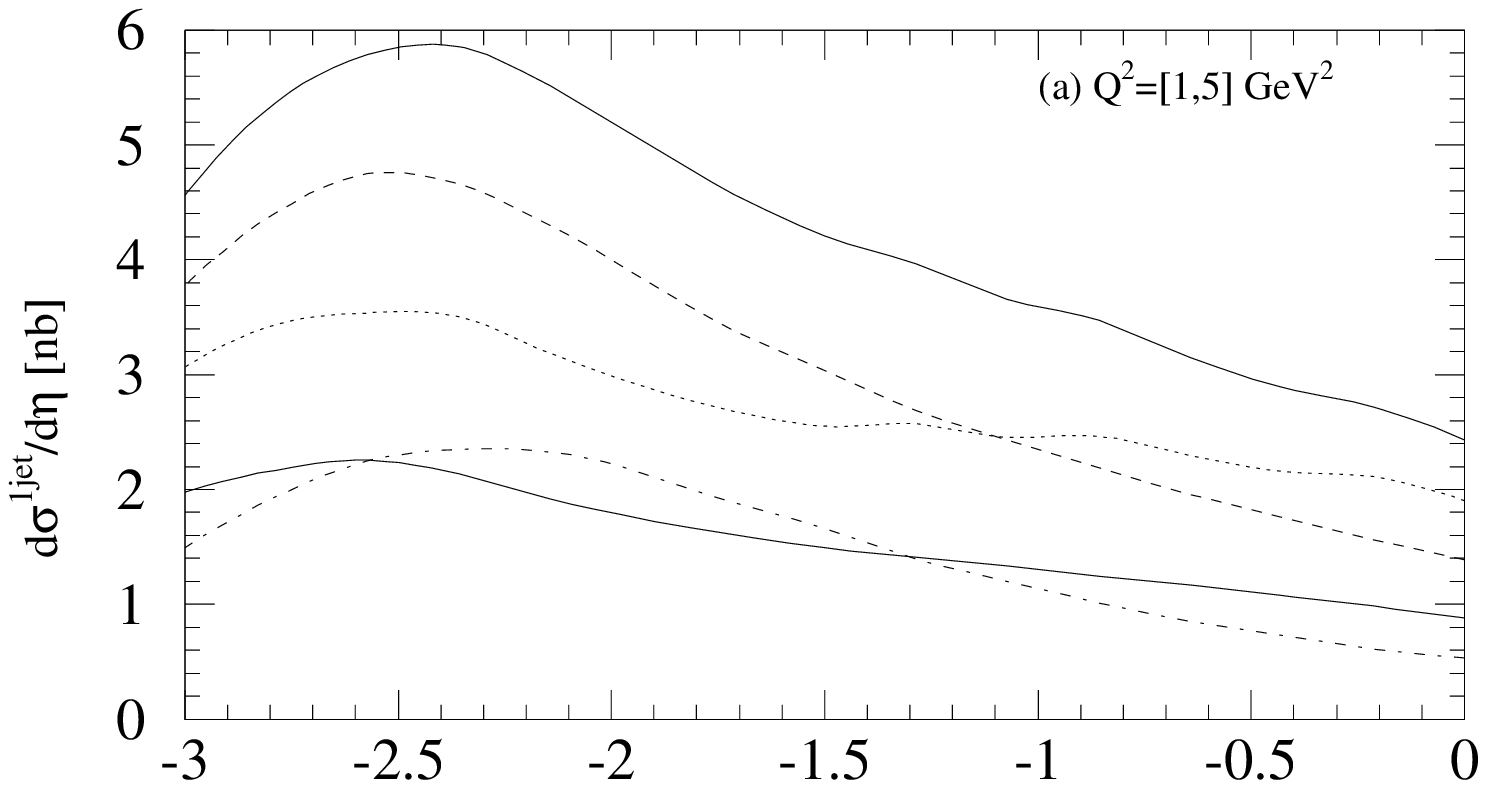,width=6.5cm,height=8cm}}
    \put(60,-2){\epsfig{file=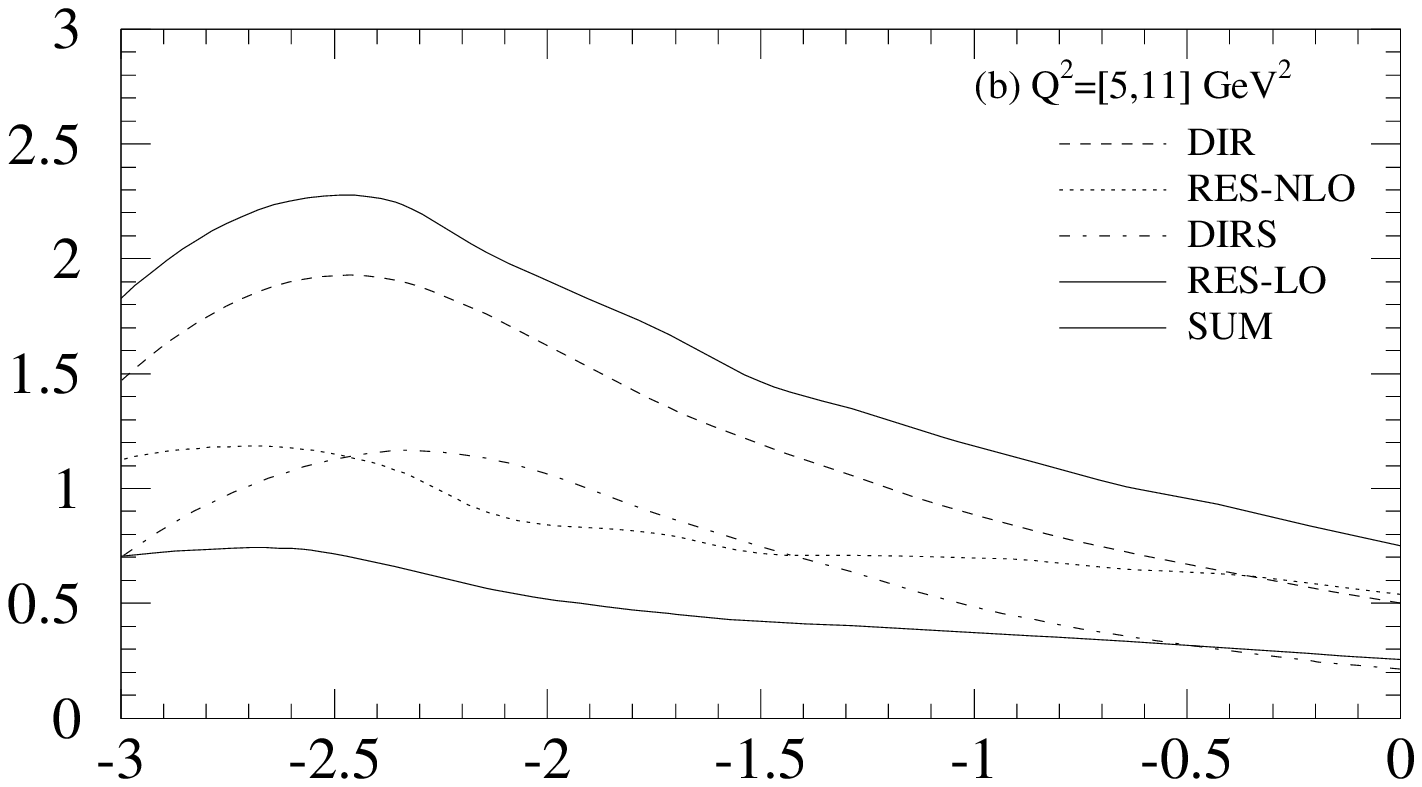,width=6.5cm,height=8cm}}
    \put(0,-37){\epsfig{file=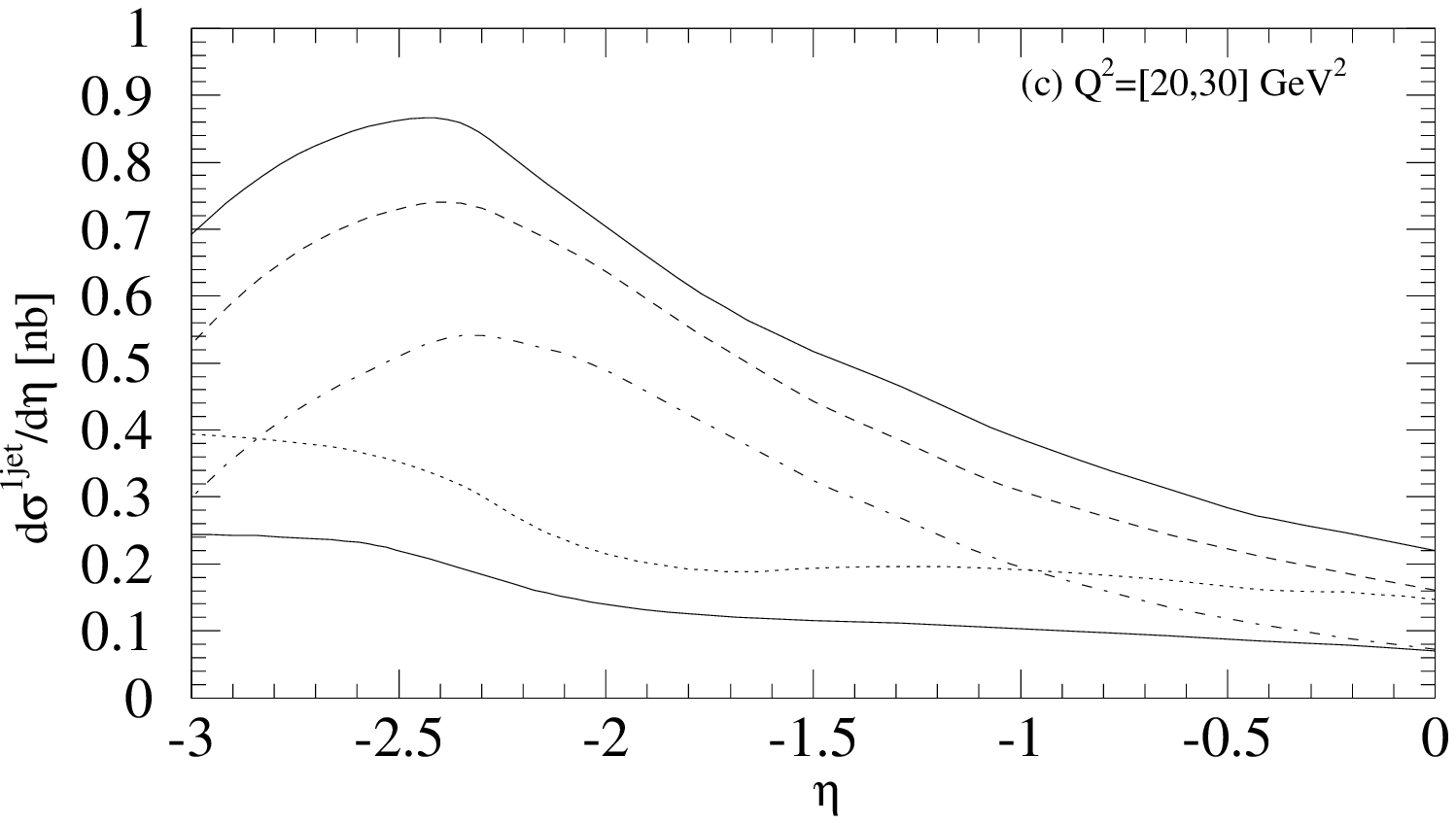,width=6.5cm,height=8cm}}
    \put(60,-37){\epsfig{file=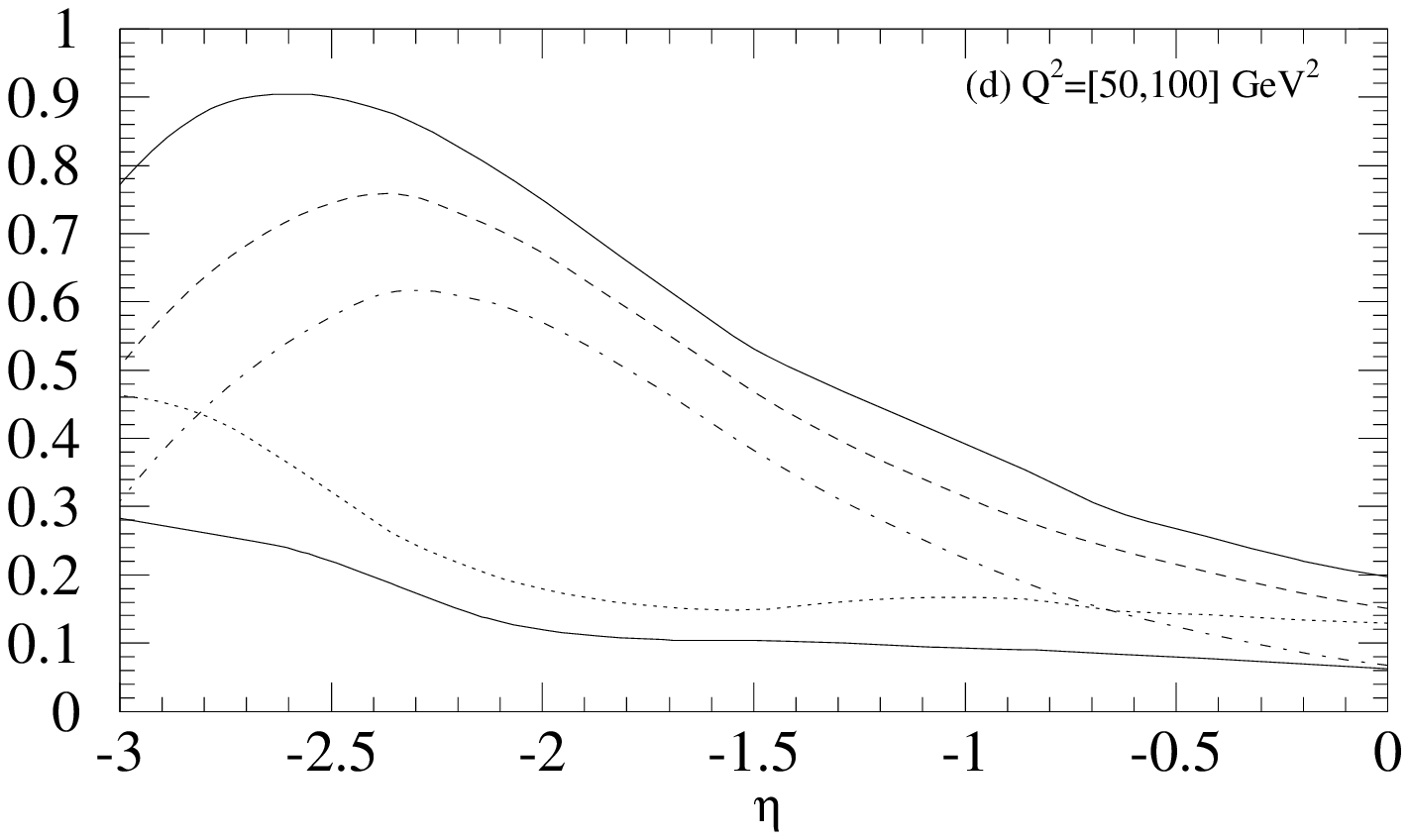,width=6.5cm,height=8cm}}
    \put(0,-2){\parbox[t]{4.7in}{\sloppy Figure 3 a-d: Inclusive
        single-jet cross section $d\sigma^{1jet}/d\eta$ integrated over
        $E_T>5$ GeV as a function of $\eta$ for four of the bins given
    in table \ref{tab}.\cite{10}}}
  \end{picture}
\end{figure}

\clearpage

\begin{table}[hhh]
\caption{Considered $Q^2$-bins covering the range of low and medium photon
  virtuality.\label{tab}}
\begin{center}
  \footnotesize
  \begin{tabular}{|c|c|c|c|c|c|c|c|} \hline
    Bin number & I & II & III & IV & V & VI & VII \\ \hline 
    $Q^2$-range in GeV$^2$ & $[1,5]$ & $[5,11]$ & $[11,15]$ & $[15,20]$ &
    $[20,30]$  & $[30,50]$ & $[50,100]$ \\ \hline
  \end{tabular}
\end{center}
\end{table}

In Fig.~3 we show single-jet inclusive $\eta$ spectra for the bins I,
II, IV and VII in the hadronic c.m.s., integrated over $E_T>5$ GeV
with $y\in [0.05,0.6]$. We use the CTEQ4M proton PDF\cite{xx} and the SaS
photon PDF.\cite{7} We show the two components RES and DIRS and
their sum (SUM, upper full curve) which should be compared to the
deep-inelastic result DIR. In addition to the NLO resolved, we have
plotted the LO resolved (lower full curve)
component. For all four $Q^2$ bins the DIR cross section in always
smaller than the cross section obtained from the sum 
of DIRS and the NLO resolved cross section.  This difference originates
essentially from the NLO corrections to the resolved cross section, as is
obvious when we add the LO resolved curve to the DIRS contribution in
Figs. 3 a--d. The sum of these agrees quite well with the DIR
curve (for a detailed discussion see \cite{10}).

\begin{figure}[bbb]
  \unitlength1mm
  \begin{picture}(80,38)
    \put(0,-33){\epsfig{file=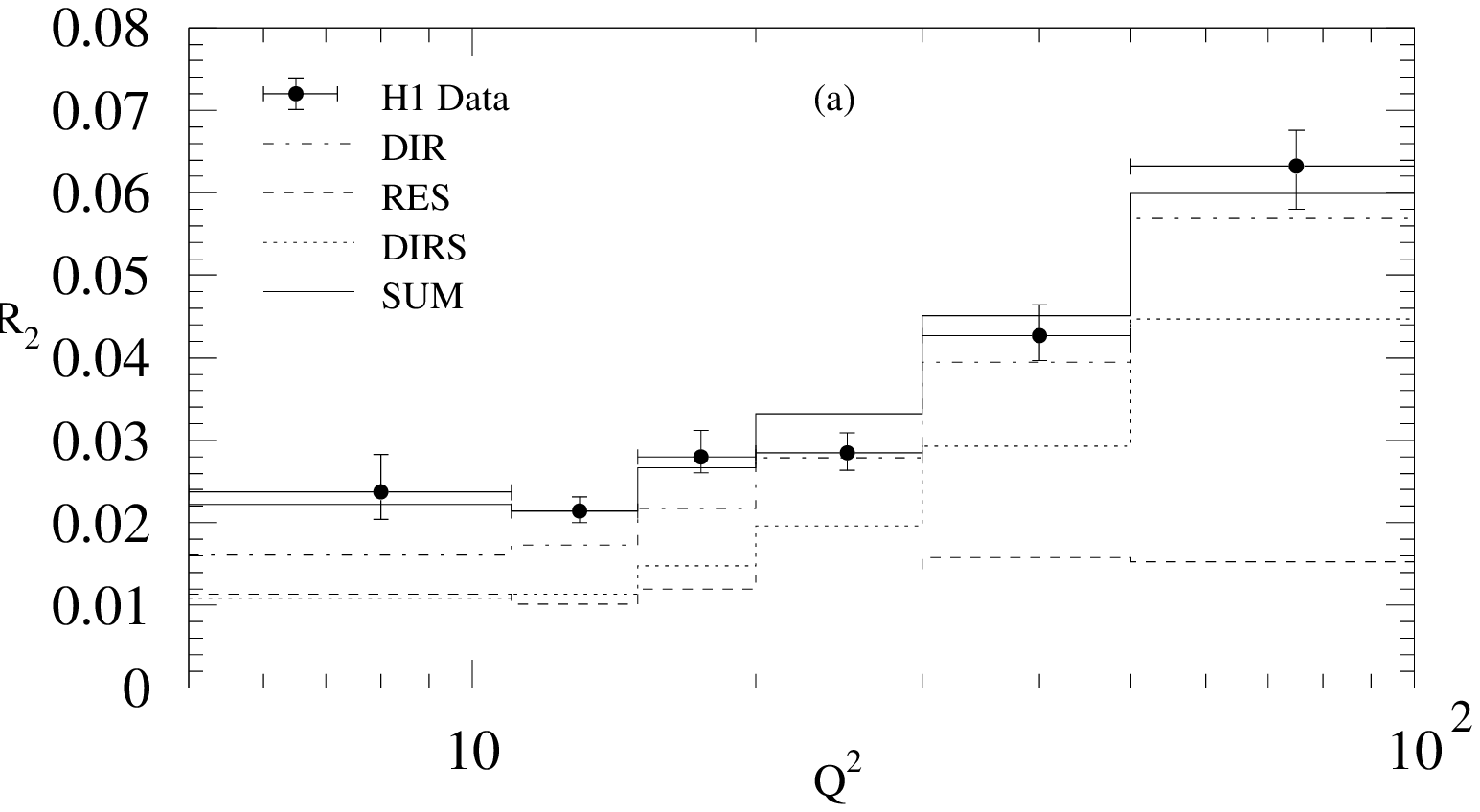,width=6.5cm,height=8cm}}
    \put(60,-33){\epsfig{file=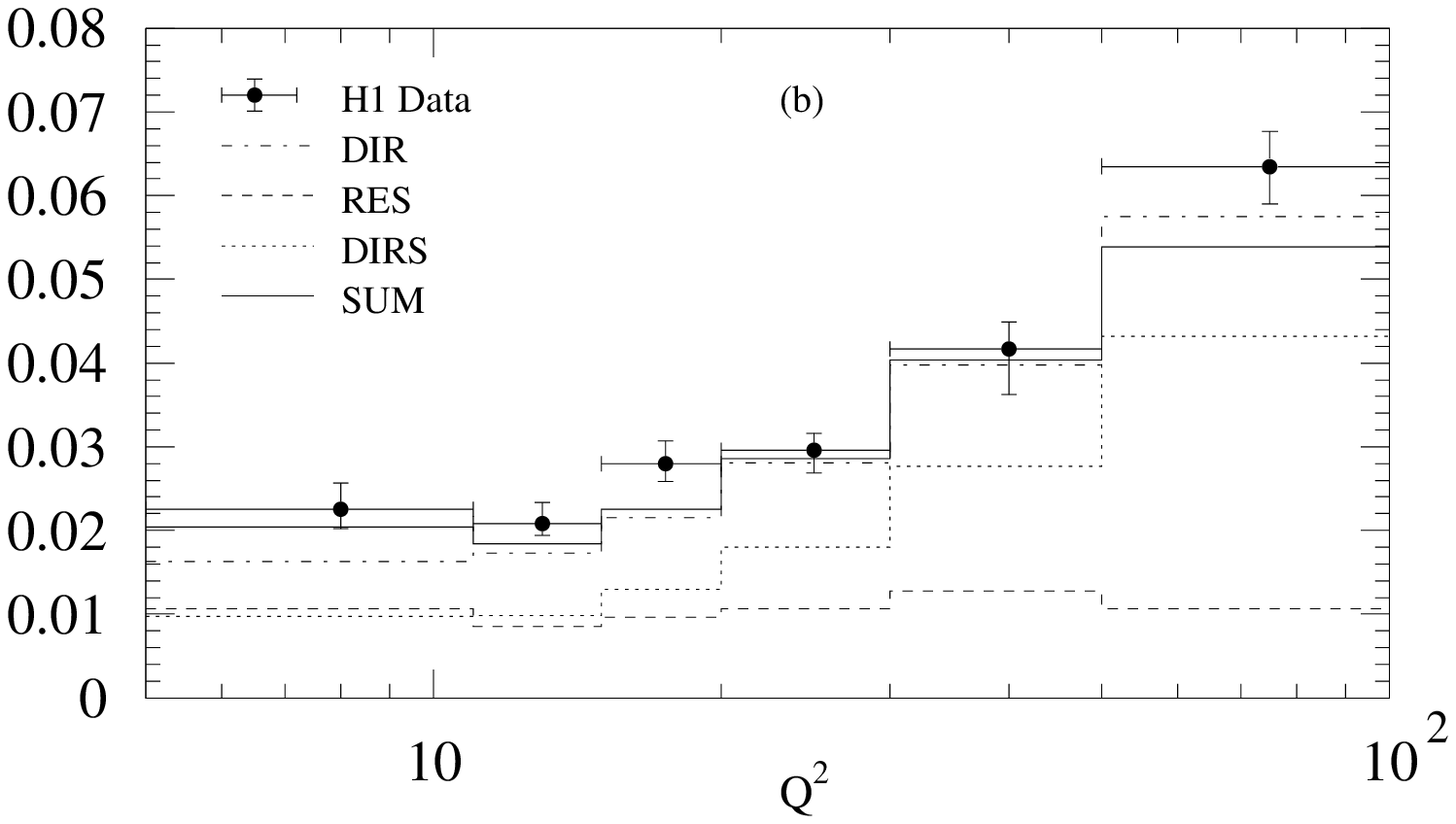,width=6.5cm,height=8cm}}
    \put(0,2){\parbox[t]{4.7in}{\sloppy Figure 4: Dijet rate $R_2$ as
        a function of $Q^2$ measured by H1,\cite{11} compared to NLO
        predictions by {\tt JetViP}. (a) $\Delta$-mode; (b)
        $\Sigma$-mode.\cite{10}}}
  \end{picture}
\end{figure}

Next we consider the exclusive dijet rate $R_2$, which measures the
cross section for two-jet production normalized to the total $ep$
scattering cross section, as a function of $Q^2$, as measured by H1.\cite{11}
The data were obtained in the bins II to VII by requiring for both
jets $E_T > 5$~GeV in the hadronic c.m.s. (the specific cuts and
details of the jet finding can be found in \cite{11}). Symmetric
cuts $E_{T_1},E_{T_2} \geq 5~GeV$ are problematic from the theoretical 
viewpoint since the so defined cross section is infrared
sensitive. With symmetric cuts on the $E_T$'s there remains no
transverse energy of the third jet, so that there is very little or no
contribution from the three-body processes. In order to avoid this
sensitivity one needs additional constraints on the $E_T$'s.
This problem has already
been discussed in connection with photoproduction.\cite{13,14} 
The H1 collaboration have chosen two additional constraints on the two
largest $E_T$'s: (i) $\Delta$-mode: if $E_{T_1}>E_{T_2}$
($E_{T_2}>E_{T_1}$) then $E_{T_1}>7~GeV~(E_{T_2}>7~GeV)$; (ii)
$\Sigma$-mode: $E_{T_1}+E_{T_2} > 13~GeV$. In Fig. 4 a and b we show
our result in the two modes for the DIR and the DIRS and RES
components for the six $Q^2$ bins II--VII. In the first $Q^2$ bin the
SUM of DIRS and RES is $50\%$ larger than the DIR (DIS)
cross section. In the smaller $Q^2$ bins SUM agrees better with the
experimental data\cite{11} than the DIR cross section. In the larger
$Q^2$ bins the difference of the cross sections DIR and SUM is small
and it can not be decided which of these cross sections agrees better
with the data due to the experimental errors.

\section*{Acknowledgments}

It is a pleasure to thank G.~Kramer for his collaboration. Useful
discussions with G.~Grindhammer, H.~Jung, H.~K\"uster and M.~Wobisch are
acknowledged. This work was supported by Bundesministerium f\"ur
Forschung und Technologie, Bonn, Germany, under Contract
05~7~HH~92P~(0), and by EEC Program {\it Human Capital and Mobility}
through Network {\it Physics at High Energy Colliders} under Contract
CHRX--CT93--0357 (DG12 COMA).
